\documentclass[amssymb,11pt,epsf]{article}
\usepackage{graphicx}
\usepackage{amssymb}

\begin{document}
\author{{\sc Dom\`enec Espriu}\footnote{espriu@ecm.ub.es} and 
{\sc Luca Tagliacozzo}\footnote{luca@ecm.ub.es}\\
Departament d'Estructura i Constituents de la Mat\`eria,
Universitat de Barcelona\\
Diagonal, 647, 08028 Barcelona, Spain}
\date{}
\title{\bf Compact lattice $U(1)$ and
Seiberg-Witten duality }
\maketitle

\bigskip\bigskip
\begin{abstract}
Simulations in compact U(1) lattice gauge theory in 4D show now beyond any
reasonable doubts that the  
phase transition  separating the Coulomb from the
confined phase is of first order, albeit a very weak one. This settles
the issue from the numerical side. On the analytical side, it was
suggested some time ago, based on  the qualitative analogy between the 
phase diagram of such a  model and the one of scalar QED  obtained by 
soft breaking the $N=2$ Seiberg-Witten model down to $N=0$,
that the phase transition should be
of second order. In this work we take a fresh look at this issue and show that 
a proper implementation of the Seiberg-Witten model below the supersymmetry 
breaking scale requires considering some
new radiative corrections. Through 
the Coleman-Weinberg mechanism this turns the second order transition
into a weakly first order one, in agreement with the numerical
results. We comment on several other aspects of this continuum model.
\end{abstract}

\vfill
\vbox{
UB-ECM-PF 02/27\null\par
December 2002}

\clearpage

\section{Introduction}

$U(1)$ pure gauge theory in four dimensions is an interacting theory when
formulated on the lattice. In fact there are two physically distinct
phases: one characterized by massless photons and long-range
interactions (the regime we would associate to continuum QED) and 
a strongly interacting region, where electrical charges are confined
and monopoles (that have a finite energy thanks to the finite value
of the lattice spacing) proliferate. It is {\em a priori} 
unclear which ``continuum'' theory is adequate to describe this regime.

A long-standing controversy regarding the order of the phase
transition separating both regimes existed for some time. Traditionally
the phase transition was believed to be of first order~\cite{Jersak:1983yz}, but
this belief was questioned in a series of papers~\cite{Jersak:1996mn}
where some evidence was provided to suggest a second order transition.
These authors  believed that 
the apparent first order nature of the phase transition was  due to
some specific configurations that on a toroidal topology  would lead to
long-lived metastable states and these  could be mistaken as footprints
of a first order transition. To avoid metastability the authors
performed a simulation using spherical topology.
The same  authors also performed
a number of interesting measurements regarding the spectrum of the
theory~\cite{Cox:1997wd}.  

Soon afterwards, it was realized that, when going to larger lattices,
a signal for a first order transition was seen also on spherical 
topologies~\cite{Campos:1998ds}, thus concluding 
that the transition was  first
order even if  a very weak one.
This seems to be the commonly accepted lore~\cite{Arnold:2002jk}. 
 
Clearly in all the above discussion an analytical approach is
lacking. It would be highly desirable to gain
some understanding of the mechanism of monopole condensation in this
specific problem, but until very recently it was unclear how to 
treat these topological excitations. Furthermore, the fact that
the transition was believed to be of first order would suggest that
indeed there was no continuum theory associated to it. 
The suggestion that the transition could be of second order, combined
with the results on $N=2$ and $N=1$ gauge theories obtained by Seiberg
and Witten~\cite{Seiberg:1994rs} led a group of authors to propose a continuum
model~\cite{Ambjorn:1997cq} for it  and, in fact, for the 
strongly coupled regime
of compact lattice $U(1)$.
 
We  shortly summarize the scenario and results presented
in ~\cite{Ambjorn:1997cq}. We shall refer in what follows 
exclusively to the continuum
model. The  starting point is $N=2$ $SU(2)$ Yang-Mills theory, 
whose action is
\begin{equation}
S_{bare}=\int d^4 x \left( \frac{1}{32\pi}Im\left( \tau \int d^2 \theta Tr W^{\alpha}W_{\alpha}\right)
+\int d^2 \theta d^2 \bar{\theta} Tr \Phi^{\dagger} e^{2gV} \Phi\right).
\end{equation}
Its  scalar potential has a flat direction and
in order to define the theory a complex parameter, the
vacuum expectation value of the scalar field, has to be fixed,
 $\langle \phi \rangle =\frac{1}{2} a\tau^3$.

At scales $\mu^2 \ll u\equiv\frac{1}{2} a^2$ the charged fields decouple and the model
is effectively described by a $N=2$, $U(1)$ gauge theory
\begin{equation}
S^{N=2}_{SW}=\frac{1}{4\pi} Im\left[ \int d^4 x \left( \int
d^{2}\theta d^{2}\bar{\theta}
\frac{\partial F}{\partial A}A^{\dagger }+\frac{1}{2}\int  d^{2}\theta \frac{\partial^{2}F}
{\partial A^{2}}W^{\alpha}W_{\alpha }\right) \right]+ {\cal O}(\frac{p^2}{\Lambda^2}),
 \label{sw}
\end{equation}
where $F$ is a known function of $A$ (the chiral superfield in the $N=2$ multiplet of the photon) and the dynamically generated
scale $\Lambda$, $W^{\alpha}$ is the Abelian field strength (for details and  notation we refer to
\cite{Seiberg:1994rs,Alvarez-Gaume:1996gd}).
For scales $\mu^2 \ge u$  the charged fields are not decoupled and they
contribute to the running. At $\mu^2\simeq u$, when the charged fields
decouple, the effective coupling constant, which is 
related to the function $F$ by $4\pi/g^2= Im\frac{\partial^2 F}{\partial a^2}$, freezes 
at the value $g=g(u)$ since below $\mu^2=u$ no fields can contribute
to its running. If $u\gg \Lambda$, this value can be safely computed 
within perturbation theory by using the beta function of $N=2$, 
$SU(2)$ Yang-Mills theory\cite{Shifman:1986vu}. The value  
$g=g(u)$ is thus the 
appropriate one to use in the effective potential at $p^2\to 0$.

As $u$ decreases and approaches $\Lambda$, perturbation theory becomes
 unreliable. The appropriate
variables to describe the theory are not $g$ and $a$, but the 
dual~\cite{Montonen:1977sn} variables $g_D$
and $a_D$, defined through the relations
\begin{equation}
a_D=\frac{\partial F}{\partial a}\qquad \frac{4\pi}{g_D^2}=Im\frac{\partial^2 F}{\partial a_D^2}=b_{11}
\end{equation}
Holomorphy and monodromy dictate that $g_D$ runs around $g_D\sim 0$
  in a way that shows that a massless hypermultiplet is present in the
spectrum: it  corresponds to a magnetic monopole.  The appropriate value of the coupling constant for long-distance
physics ($p^2\to 0$) in this  regime  is $g_D(a_D)$. This is  exactly for the same reasons that
were described above phrased this time  in term of dual variables.

The appropriate effective theory description for small 
values of $a_D$ is thus provided by~\cite{Alvarez-Gaume:1996gd}  
$S= S_D^{N=2}+S_M$, where
\begin{eqnarray} 
S_D^{N=2}&=&S_{SW}(A \rightarrow A_D,V \rightarrow V_D)                ,\cr
S_M&=&\int d^4 x \int d^4 \theta \left( M^* e^{2V_D}M+\tilde{M}^*e^{-2V_D}\tilde{M}\right)
+\int d^2 \theta \sqrt{2} A_D M \tilde{M}+ h.c.
\end{eqnarray}
and where $M$ and $\tilde{M}$ describe the $N=2$ monopole hypermultiplet.
Corrections to the above effective action will be of order $p^2/\Lambda^2$.

So far this seems to have little to do with the compact lattice $U(1)$
theory, except for the fact that the theory has a manifest $U(1)$
symmetry and that massless monopoles appear in a particular point 
of the moduli space of the theory. Clearly we need two more
ingredients: we need to enlarge the region where light monopoles appear in  
order to get a finite density of monopoles $\langle m\rangle \neq
0$ and, eventually, to break $N=2$ down to $N=0$. 
 
In~\cite{Ambjorn:1997cq} supersymmetry is broken in two
steps. First one adds a coupling between the  
chiral multiplet $\Phi$ (a member of the $N=2$ hypermultiplet) and a $N=1$
chiral superfield $z$; as a consequence the model has  now  residual
$N=1$  only. The part of the action containing the new superfield is:
\begin{equation}
S_{z}=\int d^4 x\int d^{2}\theta
d^{2}\bar{\theta}z^{\dagger }z+\left( \int d^4 x\int d^{2}\theta
lz\left( w-Tr(\Phi^2)\right) +h.c\right).  \label{dom}
\end{equation}
It introduces two free parameters $l$ and $w$.
% and allows to select  a point on the moduli space close to the value of the $w$
%parameter (so to  force the scalar component in the $N=2$  multiplet 
%of the dual photon to have a specific v.e.v.). 
This term has the net effect of enlarging the monopole
condensation region.

The limit  $l\rightarrow \infty$ used in~\cite{Ambjorn:1997cq} to
study the vacuum structure  allows to integrate the $z$ super field
out using its classical equations of motion. Unlike the original
mechanism of Seiberg and Witten~\cite{Seiberg:1994rs} the breaking introduced by 
the term $S_z$ is a hard one\footnote{It does not appear possible to 
enlarge the monopole condensation region with a soft breaking.}
 and  introduces some non-trivial quantum corrections. The obvious main effect of adding 
$S_z$ to the action is to lift the degeneracy of the vacuum so $u=\frac{1}{2}a^2$
is no longer a free parameter; in fact when taking the limit $l\rightarrow \infty$ it will be kept close  to $w$.
Again, if $u$ (now $w$) is large, perturbation theory is valid and 
a semiclassical calculation makes sense. The coupling constant now  will run with
the perturbative beta functions of $N=1$ (extended with the new matter
multiplet $z$ in the case of $l\rightarrow 0 $). However at some point, exactly as for $N=2$, the
perturbative procedure will break down. 

Then to go from $ N=1$ to $N=0$ and make contact with
the real world, the technique  discussed 
in~\cite{Alvarez-Gaume:1996gd} was used. This  consists in  coupling 
to the original $N=2$ superfield a further  $N=2$ superfield (spurion).
  The action becomes
\begin{equation}
S^{N=0}_{SW}=\frac{1}{4\pi}{Im}\left[ \int d^4 x\int
d^{2}\theta d^{2}\bar{\theta}
\frac{\partial F}{\partial A^{i}}\bar{A}^{i}+\frac{1}{2}\int d^4 x
\int d^{2}\theta 
\frac{\partial ^{2}F}{\partial A_{i}\partial A_{j}}W^{i\alpha
}W_{\alpha}^{j}
\right],\qquad  i,j=0,1.
\end{equation}
The physical fields are understood to be those labeled `1'. Breaking
to $N=0$ can be achieved by giving non-zero values either to  the $D$
or $F$ terms of the spurion (labeled `0'). In
~\cite{Ambjorn:1997cq},
 the choice of  the auxiliary fields to break
to $N=0$ is $D_{0}\neq 0, F_{0}=0 $ for the reasons mentioned there.
The starting point is thus finally $S_z+S^{N=0}_{SW}$

We are now in a position to make contact with compact lattice $U(1)$.
Let us first summarize the parameters that we have at our disposal.
If we forget about the parameter $l$ (we simply assume that it is
large enough to enforce the constraint $u\sim w$), we have two
dimensional
parameters, namely $w$ and $D_0$  and a dimensionless one, $g$, that
can be traded by $\Lambda$. If we are able to place ourselves
in a region where the characteristic momenta is much below both $w$
and $D_0$ it is clear that we will be dealing with an effective $N=0$,
$U(1)$ gauge theory. We still have some freedom in adjusting the 
relative values of $w$ and $D_0$; if by doing so we are able to 
trigger monopole condensation we shall be describing the continuum
version of the confinement-Coulomb phase transition seen in compact
lattice $U(1)$. If, in addition, we manage to do all the previous 
manipulations in a controlled manner some quantitative predictions
can be made. It is our purpose to convince the reader that this is
the case.

\section{The necessity of new quantum corrections}
When constructing supersymmetric effective actions, the 
non-renormalization
theorems\cite{Seiberg:1993vc} determine which 
perturbative and non perturbative quantum corrections are possible.
The class of possible counter terms is greatly restricted when 
supersymmetry is broken by soft terms, like in the original 
proposal to go from $N=2$ to $N=1$ of Seiberg and Witten\cite{Seiberg:1994rs} .
 
In~\cite{Alvarez-Gaume:1996gd}  $N=2$ is broken down to $N=0$ 
by the spurion mechanism allowing for a determination of the effective 
potential of the softly broken theory. (In fact, 
the first term of an 
expansion in powers of $\frac{F_0^2}{\Lambda^2}$ around the unbroken
solution is found.\footnote{As usual, terms of order 
$p^2/\Lambda^2$ are neglected.})

The scenario studied in~\cite{Ambjorn:1997cq} is somewhat
different. There the $N=2$ supersymmetry of the Seiberg-Witten model is broken
down to $N=0$  by a combination of the
 spurion mechanism we have just discussed and
a $N=1$ hard breaking term.  
This one ---unlike the 
breaking of Seiberg-Witten or the spurion mechanism--- changes the monodromy
of the original $N=2$ theory non-trivially. 

This issue was discussed in great detail in~\cite{Ambjorn:1997cq}. By
considering the $U(1)_A\times U(1)_R$ charges, it is possible to see
(modulo some highly plausible assumptions) that the requirement of
$N=1$  suffices to constrain the effective action in a way that is  sufficient for our purposes.

Since we are concerned about the monopole
condensation mechanism in compact $U(1)$ we must be in a region   
of moduli space close to $\Lambda$.
 This requires, as explained in~\cite{Alvarez-Gaume:1996gd}, the use of the
dual version of the Seiberg-Witten effective action $S_{SW}$
completed with a term for the monopole and the dualization of $S_z$,
\begin{equation}
S=S_D^{N=0} + S_M +S_z^D,\label{starting}
\end{equation}
where 
\begin{equation}
S_D^{N=0}=S_{SW}^{N=0}(A \rightarrow A_D,V \rightarrow V_D),
\end{equation}
\begin{equation}
S_z^D=\int d^4 x \left(\int d^4 \theta K(z,z^{\dagger})\int d^2 \theta l z \left\{w-U(A_D)+\frac{\Lambda^4 zl}{w}f\left( \frac{A_D^2}{w},\frac{\Lambda^4 }{w^2}\right) \right\}+h.c.\right)
\end{equation}
This involves an unknown function as well as an undetermined
K\"ahler potential
(originated by  the quantum effects of the $N=1$ term arising  when constructing the effective action~\cite{Ambjorn:1997cq}) which are  not calculable by symmetry considerations . We shall assume that the net effect of
this term when computing the scalar potential is to adjust the 
value of $a_D$; namely we exchange $w$ by $a_D$ as a free adjustable
parameter.

This, in our understanding, does not exhaust the quantum corrections we have to
include. At this point we depart from the analysis done 
in~\cite{Ambjorn:1997cq}. Let us see why the analysis presented in 
\cite{Ambjorn:1997cq} is incomplete.

We have been quite careful about the actual
meaning
of the coupling constants appearing in the effective action. They are constants
renormalized at the scale $a_D$, when the dual variables are used.
This is fine as long as the coupling constants and parameters in the
effective Lagrangian do not run from the scale $a_D$ to
$p^2=0$ where the effective potential is defined. 
This is the case in the $N=2$ theory, or  even if the theory is broken
to $N=1$ or $N=0$ when
the breaking occurs at a scale well below $a_D$. In the present case
supersymmetry is severely broken (i.e. $p^2\ll D_0  \sim a_D $)    
and the running  needs to be taken into account. This
makes the effective potential subject to radiative corrections.

We argue, in fact, that starting from energies around  $D_0$ new
quantum effects  (the standard  ones for non supersymmetric QFT)
have to be included.
To be more precise, the hierarchy of scales we have  is 
\begin{equation}
0 \leqslant  p^2\ll  b_{01}D_0\sim a_D \ll \Lambda.
\label{range}
\end{equation}
Starting from the left, the first inequality  is dictated by the
interest in studying  the non supersymmetric regime, whereas the
second allows us   to use the results 
of~\cite{Alvarez-Gaume:1996gd}.
In fact we will  use  the analytical solution of Seiberg-Witten and
its generalization to $N=0$ in (\ref{starting}) as the 
Landau-Ginzburg (tree level) approximation to the  complete effective action .
From there we still need to  scale down to zero momenta to read out  the effective potential. By doing so we shall 
show how the agreement between the 
conjecture in~\cite{Ambjorn:1997cq} and the lattice results is recovered.
In fact, the Coleman Weinberg mechanism takes place: radiative 
corrections transform a second order phase transition into a first
order one~\cite{Coleman:1973jx} .

\section{The classical vacuum structure and mass spectrum}
After elimination of the auxiliary fields and the $z$ superfield 
(through the limit $l\to \infty$)
we are left with a scalar potential that, up to constant terms, reads
\begin{equation}
V=\frac{1}{2{b}_{11}}\left( \tilde{m}^{\dagger }\tilde{m}-m^{\dagger
}m\right) ^{2}+\frac{2}{{b}_{11}}\left| a_{D}\right| ^{2}\left( mm^{\dagger
}+\tilde{m}\tilde{m}^{\dagger }\right) +\frac{{b}_{01}}{{b}_{11}}D_{0}\left(
mm^{\dagger }-\tilde{m}^{\dagger }\tilde{m}\right)
\label{treepot}
\end{equation}
In this expression $b_{01}$ and $b_{00}$ have to be understood as functions of
the point of the moduli (their analytical expressions can be 
found in~\cite{Alvarez-Gaume:1996gd}) and  $b_{11}$ is the gauge coupling (derived in~\cite{Seiberg:1994rs}). The different supersymmetry  breaking terms do
not alter  the dependence of $b_{ij}$ on $a_D$ as shown in~\cite{Ambjorn:1997cq} above the scale $D_0$. 
As discussed in the previous section, we  take
the effective potential (\ref{treepot}) as defined  at the scale $a_D$.
In (\ref{treepot}) the quantities a$_{D}$ and D$_{0}$ are free parameters.

After rotating the fields
using the $U(1)\otimes U(1)$ rigid symmetry of the above potential, it
depends only on two real
fields $m$ and $\tilde{m}$. The potential  can have, depending on the values of the parameters,
three different minima~\cite{Ambjorn:1997cq}:
$
m=\tilde{m}=0
$ 
when 
$
-2\left| a_{D}\right| ^{2}<{b}_{01}D_{0}<2\left| a_{D}\right| ^{2}
$, or
$m=0$ and $
\tilde{m}^{2}=-\left( 2{b}_{11}\left( z_{1}^{2}+z_{2}^{2}\right) -{b}%
_{01}D_{0}\right) $
when $ {b}_{01}D_{0}>2\left| a_{D}\right| ^{2}$,  
or
$\tilde{m}=0$ 
and $m^{2}=-\left( 2\left| a_{D}\right| ^{2}+{b}_{01}D_{0}\right) $
when ${b}_{01}D_{0}<-2\left| a_{D}\right| ^{2}$.

Here we will focus on the last case (The second and the third one are
interchangeable, depending on the sign of $D_0$). 
In order to  study  the critical region we introduce as control 
parameter the combination 
\begin{equation}
\alpha =-\left( 2\left| a_{D}\right| ^{2}+{b}_{01}D_0\right) .
\label{critpara}
\end{equation}
In fact, the mass of the monopole field (responsible for the critical
behavior) is given by
\begin{equation}
M_{m}^{2}=-\frac{1}{b_{11}}\left( 2\left| a_{D}\right| ^{2}+{b}_{01}D_0\right)   \label{criticalsurf}
\end{equation}
and  is thus proportional to $\alpha$. We demand (to stay close to the
transition) $ \alpha \rightarrow 0 $, while both $D_0$ and $a_D$ are
understood to be large compared to the physical scales 
(to decouple the supersymmetric modes), but small compared to $\Lambda$ (in order to trust the Seiberg-Witten effective action).

In the region we are considering   
the only light
scalar field is the monopole. The $\tilde m$ field has a large mass
\begin{equation}
M_{\tilde{m}}^{2} =\frac{8\left| a_{D}\right| ^{2}}{b_{11}},  \label{masse}
\end{equation}
and it can be safely eliminated from the effective potential which is just
\begin{equation}
V=\frac{1}{2{b}_{11}}\left( m_{1}^{2}+m_{2}^{2}\right) ^{2}+\frac{\alpha }{%
b_{11}}\left( m_{2}^{2}+m_{1}^{2}\right)
\label{potdacorr}
\end{equation}
where we have restored the imaginary part of the monopole
field $m_2$. This is of course an ordinary $\lambda \phi^4$ potential. 

As for the fermions, we have the following non diagonal mass matrix
\begin{equation}
\frac{1}{2{b}_{11}}\left( 
\begin{array}{llll}
0 & a_{D} & 0 & i\alpha \\ 
a_{D} & 0 & \alpha & 0 \\ 
0 & \alpha & 0 & \frac{1}{16\pi }\left( \frac{\partial ^{3}\mathcal{F}}{%
\partial ^{2}a_{1}\partial a_{0}}\right) D_0 \\ 
-i\alpha & 0 & \frac{1}{16\pi }\left( \frac{\partial ^{3}\mathcal{F}}{%
\partial ^{2}a_{1}\partial a_{0}}\right) D_0 & 0
\end{array}
\right) 
\label{fermmass}
\end{equation}
where the rows and columns are ordered as $\psi _{m},\psi
_{\tilde{m}},\psi _{1,}\lambda _{1}$ (see~\cite{Alvarez-Gaume:1996gd} for notation).
This matrix can be diagonalized to obtain two physical fermions with masses close to $a_{D}$
and two with masses closed to $\frac{1}{16\pi }\left( \frac{\partial ^{3}%
\mathcal{F}}{\partial ^{2}a_{1}\partial a_{0}}\right) D_0$.
Both groups are extremely heavy and can be dropped from the effective action using their equation of motion.

\section{One-Loop effective potential and phase diagram}

Using the results of Seiberg and Witten as `boundary
conditions'  at the  scale $a_D \sim b_{01}D_0$ we are now in a position to run the effective potential
down from $\mu=a_D$ to $\mu=0$. The only light fields that are
left are as expected the dual photon and the monopole field.
This is however not a generic potential: some peculiar relations
between self and electric couplings are inherited from supersymmetry. But of
course these relations shall not be preserved as supersymmetry is
broken below $D_0\sim a_D$.

The full Lagrangian, counter terms included, will be
\begin{eqnarray}
\mathcal{L} &=&\left( \partial ^{\mu }m_{1}+\frac{1}{\sqrt{b_{11}}}%
m_{2}A^{\mu }\right) ^{2}+\left( \partial _{\mu }m_{2}-\frac{1}{\sqrt{b_{11}}%
}A_{\mu }m_{1}\right) ^{2}+\frac{1}{2}\left( \left( \partial ^{2}g^{\mu \nu}-\partial ^{\mu }\partial ^{\nu }\right) A_{\mu }A_{\nu }\right) 
\label{starting Lagrangian } \nonumber\\
&&-\frac{1}{2{b}_{11}}\left( m_{1}^{2}+m_{2}^{2}\right) ^{2}-\frac{\alpha }{%
b_{11}}\left( m_{2}^{2}+m_{1}^{2}\right) %+\frac{b_{00}}{2}D^{02} 
\\ \nonumber
&& +\frac{\delta b_{11}}{2 b^2_{11}}\left( m_{1}^{2}+m_{2}^{2}\right) ^{2}+2\delta b_{e}\left( \partial _{\mu }m_{1}%
A^{\mu}m_{2}\right) 
-2\delta b_{e}\left( \partial _{\mu }m_{2}%
A^{\mu }m_{1}\right)\\ \nonumber
&&+\frac{2\delta b_{e}}{\sqrt{b_{11}}}\left( m_{2}^{2}+m_{1}^{2}\right)A^{\mu}A_{\mu} -\frac{\delta M^2}{b_{11}} \left(
m_{2}^{2}+m_{1}^{2}\right).% +\delta b_{e}^2A_{\mu A^{\mu} \left(m_{2}^2+m_{1}^2\right). 
\end{eqnarray}
We shall work in the Landau gauge.

The one loop correction to the effective potential 
in $\overline{MS}$  produces the following effective potential
\begin{eqnarray}
V &=&\frac{1}{2{b}_{11}}m^{4}+\frac{\alpha }{b_{11}}m^{2}+\frac{1}{\left( 4\pi \right) ^{2}}\left( \frac{3}{{b}_{11}}%
m^{2}+\frac{\alpha }{b_{11}}\right) ^{2}\left( \log \frac{%
3m^{2}+\alpha }{\mu^{2}}-\frac{3}{2}\right)  \label{1looppot} \\
&&+\frac{1}{\left( 4\pi \right) ^{2}}\left( \frac{1}{{b}_{11}}m^{2}+%
\frac{\alpha }{b_{11}}\right) ^{2}\left( \log \frac{m^{2}+\alpha }{\mu^{2}%
}-\frac{3}{2}\right) +\frac{3}{\left( 4\pi \right) ^{2}}\frac{m^{4}}{%
b_{11}^{2}}\left( \log \frac{m^{2}}{\mu^{2}}-\frac{5}{6}\right)  \nonumber
\end{eqnarray}
in agreement with~\cite{Martin:2001vx}, for instance. We have used
again the residual $U(1)$ rigid symmetry to get rid of $m_2$ and 
we have replaced $m_1\to m$. Here $\mu^2$ is the subtraction scale.

From the one loop calculations we also extract the beta functions
\begin{equation}
\beta _{b_{11}}=\frac{1}{\left( 4\pi \right) ^{2}}\frac{13}{{b}_{11}^{2}}\qquad
\beta _{b_{e}}=\frac{1}{24\pi ^{2}}\left( 
\frac{2}{b_{11}}\right) ^{\frac{3}{2}},\label{beta}  
\end{equation}
which are of course different.  

With these results what follows is quite predictable.
We are interested in the phase transition from the Coulomb to the confined
phase. The order parameter is $m$.  A value of $m$  different from
zero reveals monopole condensation and the presence of the dual 
Meissner effect.

By changing $\alpha $ we move from a phase with $m\neq 0$ to one 
with $m=0$ . If we tune $\alpha $  to have zero renormalized mass 
(to stay on the critical surface) the presence of a  minimum in
the potential  outside from the origin implies a first order 
phase transition~\cite{Coleman:1973jx}. 

On the critical surface,
there are in the effective theory two distinct coupling constants (the gauge coupling $b_e$ and the quartic self coupling for the scalar field $b_{11}$) that
have the same  bare value , as dictated by supersymmetry. We call this bare value $b_{11}^{SW}$. 
In spite of this, they are different below $a_D$  as they  run  with  different beta functions (a naive inspection of the bare potential would lead to the erroneous conclusion that there is just one coupling).
The existence of two beta functions, as pointed out by Yamagishi~\cite{Yamagishi:1981qq}, is the  essential ingredient that allows,  in scalar QED, (as in other models~\cite{Espriu:1998vc}) radiative corrections to induce  a first order transition~\cite{Coleman:1973jx}. 
Following the analysis  in~\cite{Yamagishi:1981qq}  we obtain the one loop RG improved effective potential
\begin{equation}
V=\frac{b_{11}(t)}{4!} m^4 exp \left[ 4 \int_0 ^t dt' \frac{\gamma(t')}{1-\gamma(t')} \right]
\end{equation} 
where $t=ln(m/\mu)$ and $\gamma(t)=\gamma(b_{11}(t),b_e(t))$ with $b_e(t),b_{11}(t)$ solution of the differential equations
\begin{equation}
\frac{d b_{11}(t)}{d t}=\frac{\beta_{b_{11}}(b_{11},b_e)}{1-\gamma(b_{11},b_e)}; \qquad
\frac{d b_{e}(t)}{d t}=\frac{\beta_{b_{e}}(b_{11},b_e)}{1-\gamma(b_{11},b_e)} \label{trajectory}
\end{equation}
with initial conditions
\begin{equation}
b_{11}(0)=b_{11}^{SW} \qquad b_e(0)=b_{11}^{SW}
\end{equation}
The  existence of a minimum away from the origin  implies that the trajectories (\ref{trajectory}) in the $(b_{11},b_{e})$ plane should cross the line
\begin{equation}
4b_{11}+\beta_{b_{11}}=0\label{critical}.
\end{equation}
 
\begin{figure}[tbp]
\begin{center}
\includegraphics[width=10cm,angle=-90]%
{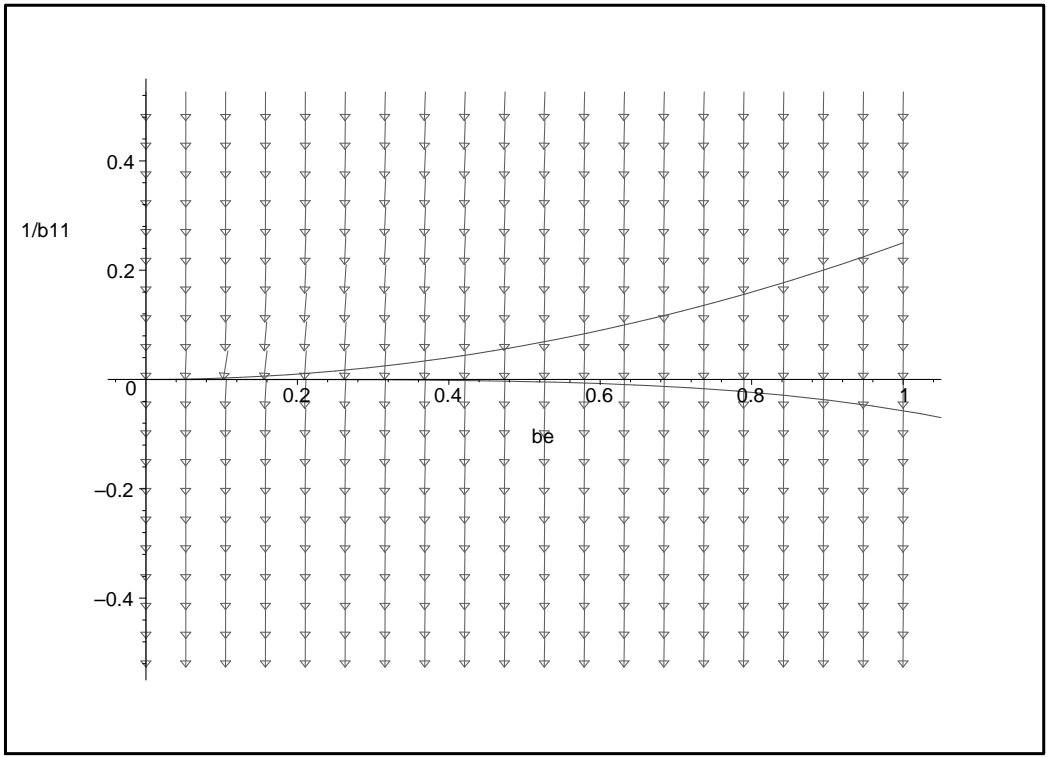}
\end{center}
\caption{Flux of the couplings along the renormalization group trajectories, $b_{11}$ versus $b_{e}$. Starting with bare conditions on the upper line the flux  always crosses the  lower line representing \ref{critical}. This is the footprint of the  first order transition}\label{flux}
\end{figure}

\begin{figure}[tbp]
\begin{center}
\includegraphics[width=10cm,angle=-90]%
{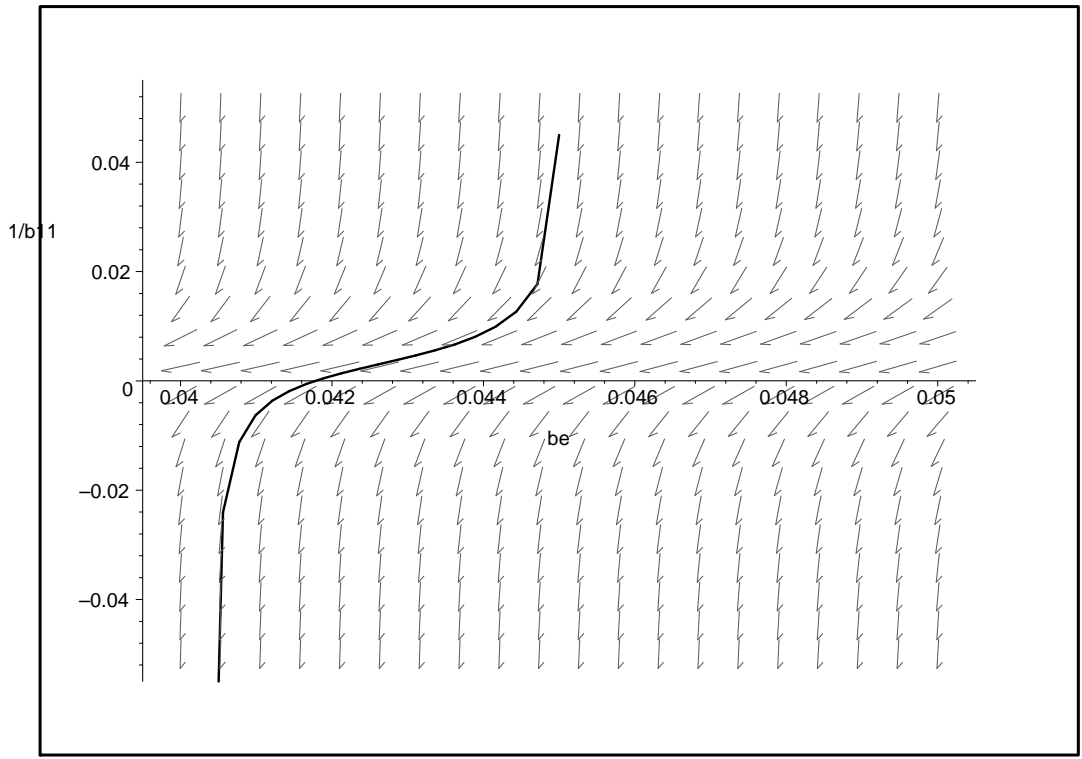}
\end{center}
\caption{Zoom of the previous plot in the perturbative region (small $b_e$). The continuous line sketches the flux of the couplings along the renormalization group trajectories in the  $b_{11}$,$b_{e}$ plane for some typical initial conditions}\label{flux2}
\end{figure}

The  plot of the renormalization-group trajectories of the coupling constants (tangent to the arrow field in figure \ref{flux}) shows  that the trajectories leaving the bare curve $\frac{1}{b_{11}}=\frac{b_e^2}{2}$ (upper solid line in figure \ref{flux})  always cross the line (\ref{critical}) (lower solid line in figure \ref{flux}).
This  implies that  the transition is of first order in all the plane but the origin. A zoom of the previous plot in the perturbative region with explicitly sketched flow is drawn in figure \ref{flux2}.

The fact that the transition is first order can also be checked  by direct inspection of the potential (\ref{1looppot}); that is  by plotting (\ref{1looppot}) for typical values of the parameters and  looking to the position of the minima.

\begin{figure}[tbp]
\begin{center}
\includegraphics[width=10cm,angle=-90]%
{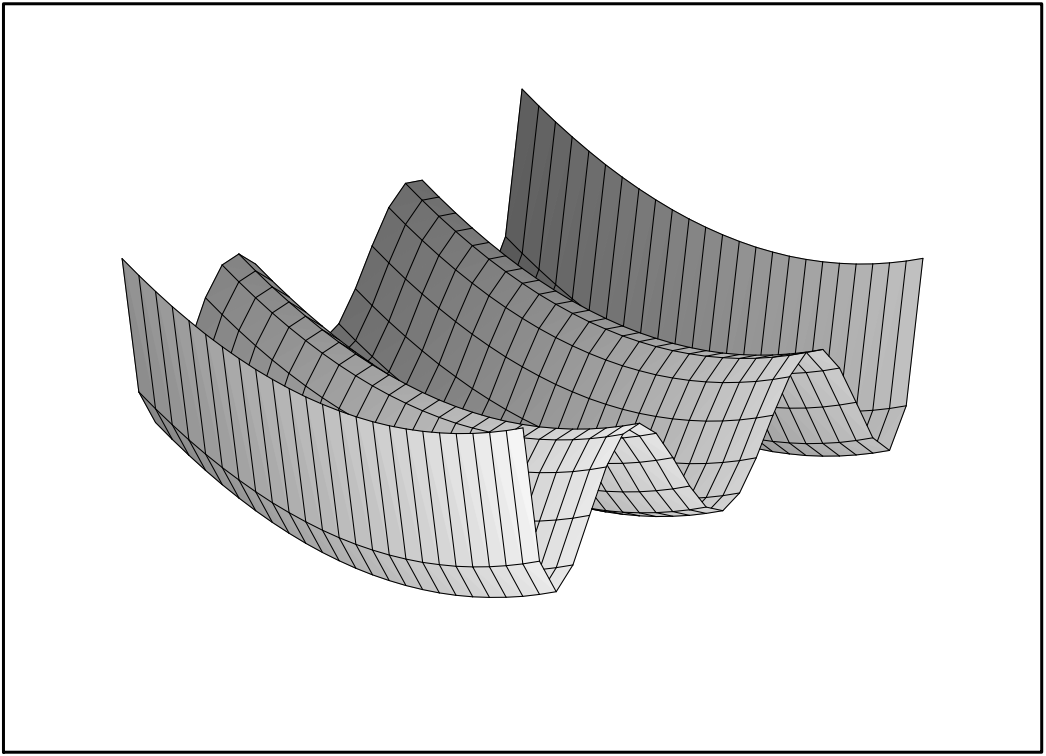}
\end{center}
\caption{First order phase transition: the potential as function of $m$ and
$\tilde{m}$ with $\alpha=0.008$ shows the coexistence of the Coulomb and confined phases.}\label{3dtransition}
\end{figure}

It  is evident  that the system passes from the Coulomb
phase  to the confined one  through a first order transition (fig. \ref{3dtransition}).
We thus conclude that the transition is indeed of first
order. However, as befits a radiatively induced first order
transition, this is expected to be quite weak, exactly as observed in lattice simulations.

\section{Conclusions}

The agreement between the predictions of Seiberg-Witten theory and
what is observed in lattice simulations of compact $U(1)$ goes 
beyond the fact that both indicate the existence of a weakly first 
order transition. As it was emphasized in \cite{Ambjorn:1997cq}, 
the parity and charge-conjugation quantum numbers of the lightest 
one-particle states in the confinement phase ($ 0^{++}$ and $1^{+-}$) seem to  agree with the clearest signal of ``gauge-ball'' state from lattice simulation~\cite{Cox:1997wd}. The fact that the transition is weekly first order makes the comparison between continuum and lattice results meaningful.

It might seem that, at the end of the day after the successive
supersymmetry breakings, very little is left of
the bare Lagrangian with $N=2$, $SU(2)$ Yang-Mills gauge invariance
and that one is left with the Lagrangian of scalar electrodynamics.
This perception is indeed correct, but it was by no means obvious
that this was the appropriate theoretical framework to describe 
the confining-Coulomb transition in compact lattice $U(1)$. It was not
even obvious that the appropriate theory could be written in terms of
local variables. The good theoretical control provided by the
Seiberg-Witten model allows us to be on firm ground.

The supersymmetry breaking is well under control. The problem we are
interested is considerably easier than trying to get predictions for,
say, $SU(2)$ gauge theory, because we are in a kinematical regime
where it is consistent to have $D_0\ll \Lambda$. As far as we can see,
the theoretical prediction is quite robust.

The results obtained here modify the conclusions of
~\cite{Ambjorn:1997cq}. Some radiative corrections, necessary below the 
supersymmetry breaking scale, were overlooked in that work. Their 
consideration reconciles the qualitative agreement between the
phase diagram of the $U(1)$ theory obtained after breaking the 
supersymmetry of the Seiberg-Witten theory and the phase diagram of
compact $U(1)$ lattice theory in four dimensions. This also supports 
the idea that confinement in four dimensions is due to monopole 
condensation as already demonstrated in the three-dimensional 
case~\cite{Polyakov:1977fu}.
We are currently considering the application of the same techniques to other models.
%\begin{figure}[p]
%\begin{center}
%\includegraphics[width=10cm,angle=-90]{diso}
%\end{center}
%\caption{Coulomb Phase: the potential as function of $m$ with  $\alpha=0.01$}\label{coulomb}
%
%\end{figure}

%\begin{figure}[p]
%\begin{center}
%\includegraphics[width=10cm,angle=-90]%
%{trans}
%\end{center}
%\caption{First order phase transition: potential as function of m$_{01}$ with $\alpha=0.008$
%coexistence of Coulomb and Confined phase}\label{transition}
%
%\end{figure}

%\begin{figure}[tbp]
%\begin{center}
%\includegraphics[width=10cm,angle=-90]%
%{ord}
%\end{center}
%\caption{Confined Phase: potential as function of m$_{01}$ with $\alpha=0.0001$}
%\label{confined}%
%\end{figure}
\begin{figure}[tbp]
\begin{center}
\includegraphics[width=10cm,angle=-90]%
{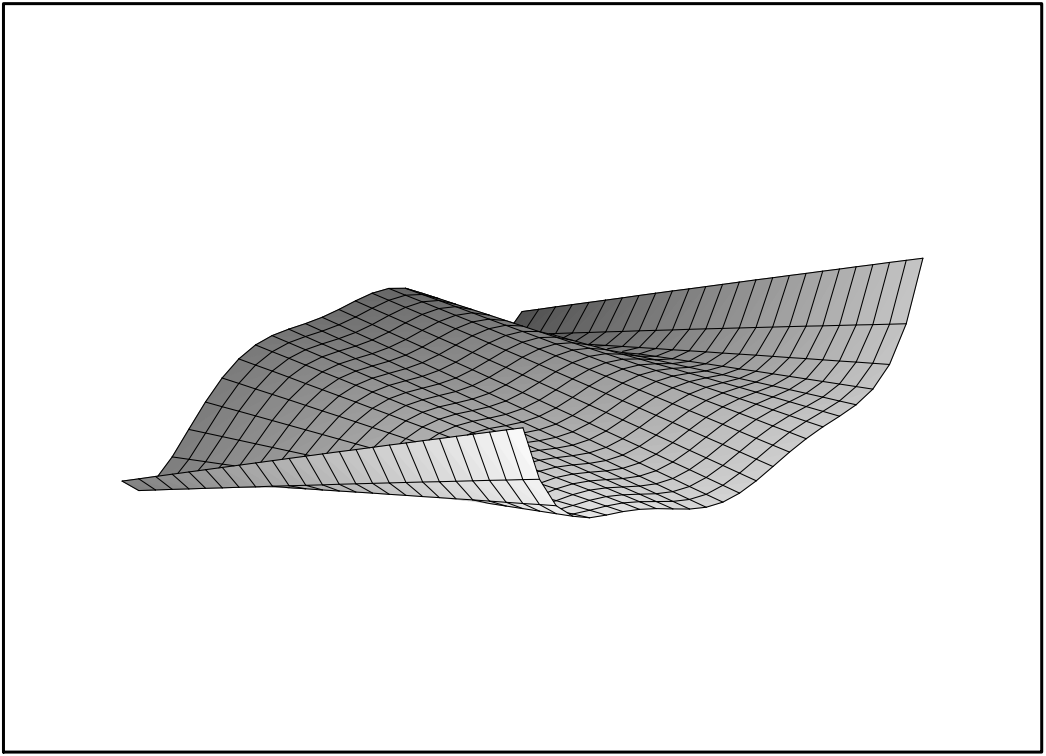}
\end{center}
\caption{Dynamical evolution: from Coulomb to confined through a first order
transition. Potential as function of $m$  and $\alpha $.}\label{dinamic}
\end{figure}

%\begin{figure}[tbp]
%\begin{center}
%\includegraphics[width=10cm,angle=-90]%
%{3ddiso}
%\end{center}
%\caption{Coulomb phase.Potential as function of m$_{01}$ and \~{m}$_{01}$
% with $\alpha=0.01$} \label{3dCoulomb}
%
%\end{figure}

%\begin{figure}[tbp]
%\begin{center}
%\includegraphics[width=10cm,angle=-90]%
%{3dord}
%\end{center}
%\caption{Confined phase. Potential as function of m$_{01}$ and \~{m}$_{01}$ with $\alpha=0.001$}\label{3dConfined}
%
%\end{figure}

%\begin{thebibliography}
%\bibliographystyle{prsty}
%\bibliography{luca}
{}

%\end{thebibliography}

\end{document}